\documentclass[aps,prd,superscriptaddress,floatfix,twocolumn,nofootinbib,preprintnumbers]{revtex4}

\usepackage{amsmath,amssymb,amsfonts,dcolumn,slashed,dsfont,multirow}
\usepackage{url}
\usepackage{graphicx}
\usepackage{psfrag}

\newcommand{\beq}{\begin{equation}}
\newcommand{\eeq}{\end{equation}}
\newcommand{\mc}{m_\chi}
\newcommand{\mN}{m_N}
\newcommand{\mA}{m_A}
\newcommand{\muN}{\mu_N}

\newcommand{\diff}{\text{d}}

\newcommand{\Op}{\mathcal{O}}
\newcommand{\qq}{\mathbf{q}}
\newcommand{\vv}{\mathbf{v}}
\newcommand{\F}{\mathcal{F}}

\providecommand{\keV}{\,\text{keV}}
\providecommand{\MeV}{\,\text{MeV}}
\providecommand{\GeV}{\,\text{GeV}}
\providecommand{\TeV}{\,\text{TeV}}

\allowdisplaybreaks[1]

\begin{document}

\preprint{INT-PUB-18-006}

\title{Discriminating WIMP--nucleus response functions in present and future\\ XENON-like direct detection experiments}

\author{A.\ Fieguth}
\email[E-mail:~]{a.fieguth@uni-muenster.de}
\affiliation{Institut f\"ur Kernphysik, Westf\"alische Wilhelms-Universit\"at M\"unster, 48149 M\"unster, Germany}
\author{M.\ Hoferichter}
\email[E-mail:~]{mhofer@uw.edu}
\affiliation{Institute for Nuclear Theory, University of Washington, Seattle, WA 98195-1550, USA}
\author{P.\ Klos}
\email[E-mail:~]{pklos@theorie.ikp.physik.tu-darmstadt.de}
\affiliation{Institut f\"ur Kernphysik, Technische Universit\"at Darmstadt, 64289 Darmstadt, Germany}
\affiliation{ExtreMe Matter Institute EMMI, GSI Helmholtzzentrum f\"ur Schwerionenforschung GmbH, 64291 Darmstadt, Germany}
\author{J.\ Men\'endez}
\email[E-mail:~]{menendez@cns.s.u-tokyo.ac.jp}
\affiliation{Center for Nuclear Study, The University of Tokyo, 113-0033 Tokyo, Japan}
\author{A.\ Schwenk}
\email[E-mail:~]{schwenk@physik.tu-darmstadt.de}
\affiliation{Institut f\"ur Kernphysik, Technische Universit\"at Darmstadt, 64289 Darmstadt, Germany}
\affiliation{ExtreMe Matter Institute EMMI, GSI Helmholtzzentrum f\"ur Schwerionenforschung GmbH, 64291 Darmstadt, Germany}
\affiliation{Max-Planck-Institut f\"ur Kernphysik, Saupfercheckweg 1, 69117 Heidelberg, Germany}
\author{C.\ Weinheimer}
\email[E-mail:~]{weinheimer@uni-muenster.de}
\affiliation{Institut f\"ur Kernphysik, Westf\"alische Wilhelms-Universit\"at M\"unster, 48149 M\"unster, Germany}

\begin{abstract}
The standard interpretation of direct-detection limits on dark matter involves particular assumptions of the underlying WIMP--nucleus interaction, such as, in the simplest case, the choice of a Helm form factor that phenomenologically describes an isoscalar spin-independent interaction.
In general, the interaction of dark matter with the target nuclei may well proceed via different mechanisms, which would lead to a different shape of the corresponding nuclear structure factors as a function of the momentum transfer $q$. We study to what extent different WIMP--nucleus responses can be differentiated based on the $q$-dependence of their structure factors (or ``form factors''). We assume an overall strength of the interaction consistent with present spin-independent limits and consider an exposure corresponding to XENON1T-like, XENONnT-like, and DARWIN-like direct detection experiments. We find that, as long as the interaction strength does not lie too much below current limits, the DARWIN settings allow a conclusive discrimination of many different response functions based on their $q$-dependence, with immediate consequences for elucidating the nature of dark matter.    
\end{abstract}

\maketitle

\section{Introduction}
\label{sec:introduction}

The search for weakly interacting massive particles (WIMPs) as candidates for dark matter is pursued with increasing effort in direct detection experiments~\cite{Baudis:2016qwx} that aim to observe hints for WIMPs scattering off target nuclei, see, e.g.,~\cite{Angloher:2015ewa,Agnese:2015nto,Agnes:2015ftt,Amole:2016pye,Armengaud:2016cvl,Akerib:2016vxi,Cui:2017nnn,Aprile:2017iyp} for recent limits.
Among the available technologies, dual-phase xenon time projection chamber (TPC) detectors lead the search for WIMP masses $\mc\gtrsim5\GeV$~\cite{Cui:2017nnn,Akerib:2016vxi,Aprile:2017iyp}. 
The third generation experiment XENON1T presented first results from 34.2 live days utilizing $\sim 1\,\text{t}$ fiducial mass~\cite{Aprile:2017iyp}, and is taking further data in order to achieve its projected sensitivity of a single-nucleon cross section of $\sigma_0=1.6\times 10^{-47}\,\text{cm}^2$ for a WIMP mass $\mc=50\GeV$ after an exposure of $2\,\text{ton}\,\text{years}$~\cite{Aprile:2015uzo}. 
Ongoing efforts are expected to push sensitivities further in the next years, including the LZ detector~\cite{Akerib:2015cja}, PandaX-xT~\cite{PandaXxT}, and an upgrade of XENON1T to the XENONnT experiment, with a planned active mass about three times larger~\cite{Aprile:2015uzo}.
On a longer time scale, the proposed DARWIN experiment~\cite{Aalbers:2016jon} aims to cover the full parameter space before the coherent neutrino--nucleus scattering~\cite{Akimov:2017ade} becomes the dominant background.

Traditionally, experimental analyses assume a standard isoscalar spin-independent (SI) interaction between the WIMP and the xenon nuclei, with the underlying nuclear structure factor (also referred to as ``form factor'') approximated by a so-called Helm form factor~\cite{Helm:1956zz}. This choice is motivated by the fact that if the standard SI isoscalar WIMP--nucleon interaction is present and not suppressed, it is expected to be the dominating contribution given the coherent enhancement by all $A$ nucleons in a target nucleus. The differential WIMP scattering rate per unit energy 
\beq
\frac{\diff R}{\diff E} \propto \frac{\diff \sigma}{\diff q^2} \propto \sigma_0 \times |\F(q^2)|^2 \,,
\label{eq:proportionality}
\eeq 
is proportional to the product of the WIMP--nucleon cross section $\sigma_0$ and the nuclear structure factor $\F$ (normalized by $\F(0)=A$ for the standard SI response).
The momentum transfer $q$ is related to the recoil energy $E$ and the mass of the xenon nucleus $\mA$ by $E=q^2/2\mA$.
The limits derived from experimental analyses are usually presented in terms of the WIMP--nucleon cross section $\sigma_0$. However, Eq.~\eqref{eq:proportionality} shows that the limit deduced on $\sigma_0$ for a given differential WIMP recoil spectrum depends on the shape of $\F$ as well as on its normalization. In particular, the standard limits on $\sigma_0$ are valid only if the SI interaction is indeed the dominating contribution to WIMP--nucleus scattering.  

However, if the isoscalar SI interaction is suppressed, e.g., in the vicinity of so-called blind spots in supersymmetric models~\cite{Cheung:2012qy,Huang:2014xua,Crivellin:2015bva}, other channels become important and can even be dominant.
This issue has been addressed experimentally by dedicated searches of alternative channels, e.g., those based on spin-dependent (SD) WIMP--nucleus interactions~\cite{Aprile:2013doa,Uchida:2014cnn,Fu:2016ega,Akerib:2017kat}, non-relativistic effective field theory (NREFT) operators~\cite{Schneck:2015eqa,Aprile:2017aas}, or generically $q$-suppressed responses~\cite{Angloher:2016jsl}.
More generally, the dynamics of the underlying strong interaction, quantum chromodynamics (QCD), as implemented in the framework of chiral effective field theory (EFT)~\cite{Epelbaum:2008ga,Machleidt:2011zz,Hammer:2012id,Bacca:2014tla} -- an effective theory of QCD valid at nuclear structure energies and momentum transfers of the order of the pion mass -- predicts further classes of subleading corrections. Some of these responses can be coherently enhanced even if the leading SI contribution vanishes~\cite{Hoferichter:2016nvd}.

In the event of a WIMP detection, the underlying WIMP--nucleus interaction encodes valuable information about the nature of the dark matter. Thus, it is natural to ask whether different interactions could be distinguished in present and future experiments, one possible strategy relying on the different $q$-dependence of the nuclear structure factors involved.
While recent work has also studied the potential discriminating power of direct detection experiments for a set of NREFT operators~\cite{Rogers:2016jrx},
in this paper we study to what extent such different interactions can actually be discriminated in a realistic experimental setting.  

To this end we consider the nuclear structure factors derived in~\cite{Hoferichter:2016nvd}, excluding all other contributions but one operator of interest, and investigate whether its different $q$-dependence allows one to distinguish it from the standard Helm form factor. Our investigation derives signal and background distributions based on toy-Monte Carlo (toy-MC) simulations, combining the parameters of the XENON100 detector with standard astrophysical assumptions~\cite{Lewin:1995rx} and the structure factor of interest. 
While we fix the parameters specific to the known parameters from the XENON100 detector (e.g., the search window in energy), we scale the total signal and background rates to the total mass and expected background of the present XENON1T, future XENONnT (also representative of LZ~\cite{Akerib:2015cja} or PandaX-xT~\cite{PandaXxT}), and proposed DARWIN experiments. For a given WIMP mass $\mc$, we set the signal strength by fixing the expected number of WIMP events corresponding to a WIMP--nucleon cross section $\sigma_0$ in the standard SI interpretation. 
Then we perform a likelihood-ratio analysis to quantify the discrimination power of each experiment for the signals caused by various non-standard WIMP--nucleus interactions with respect to the signal caused by the standard Helm form factor.

The remainder of the article is structured as follows. First, we provide a short review of the respective nuclear structure factors in Sec.~\ref{sec:theory}, before describing in more detail the detector assumptions and the statistical method in Sec.~\ref{sec:method}. Our results are presented in Sec.~\ref{sec:results}, with conclusions summarized in Sec.~\ref{sec:conclusions}.

\section{Theory}
\label{sec:theory}

The rate for a WIMP scattering elastically off a nucleus is given by
\beq
\label{rate}
\frac{\diff R}{\diff q^2}=\frac{\rho M}{\mA\mc}\int_{v_\text{min}}^{v_\text{esc}}\diff^3 v\, v f(v)\,\frac{\diff \sigma}{\diff q^2} \,,
\eeq
where $q=|\qq|$ is the three-momentum transfer between the WIMP and the nucleus, $v=|\vv|$ the velocity of the WIMP with distribution $f(v)$,
$\mc$ and $\rho$ the WIMP mass and density, respectively,
$M$ is the active mass of the experiment, and $\mA$ the mass of a target nucleus. Here we take the astrophysical parameters\footnote{Here, we simplify the notation of the velocity integral for better readability. The respective minimum velocity $v_\text{min}$ required for a certain momentum transfer $q$ has to be considered in the rest frame of the detector, i.e., the Earth, whereas the escape velocity $v_\text{esc}$ has to be treated in the rest frame of our Galaxy.} from~\cite{Lewin:1995rx} (the impact of the astrophysical uncertainties is considered, e.g., in~\cite{Kahlhoefer:2016eds,Catena:2018ywo,Krauss:2018pvg}).
Our focus lies on the cross section $\diff \sigma/\diff q^2$,
which for the leading isoscalar SI interaction reads
\beq
\label{SI_simp}
\frac{\diff \sigma}{\diff q^2}=\frac{\sigma_0}{4v^2\mu_N^2}\F_\text{SI}^2(q^2)\,,\qquad \muN=\frac{\mN\mc}{\mN+\mc}\,,
\eeq
with nucleon mass $\mN$.
In this way, limits on the WIMP scattering rate $\diff R/\diff q^2$, studied in a particular range of momentum transfer $q$, can be condensed into a limit on the WIMP--nucleon cross section $\sigma_0$.
The cross section is written in terms of a nuclear structure factor, $\F_\text{SI}(q^2)$, that is well approximated by a phenomenological Helm form factor~\cite{Helm:1956zz,Lewin:1995rx,Vietze:2014vsa}.
This form factor finds its maximum at $\F_\text{SI}(0)=A$, where all $A$ nucleons in the nucleus contribute coherently, and decreases exponentially for finite values of the momentum transfer, see Fig.~\ref{fig:structure_factors} below.
The typical scale of the relevant momentum transfers is 
\beq
q\sim \frac{m_A\mc}{m_A+\mc} v \sim 10^{-3}m_A\sim100\MeV\,.
\eeq

\begin{figure}[t] 
	\centering
	\includegraphics[width=\columnwidth,clip]{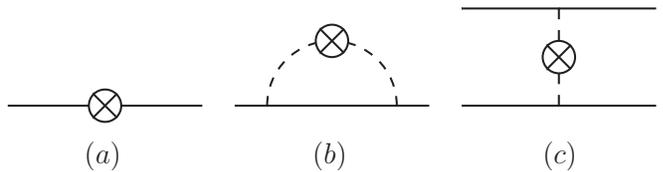}
	\caption{Diagrams for WIMP--nucleon interactions in chiral EFT. Solid (dashed)
		lines refer to nucleons (pions) and crosses indicate the coupling to the WIMP. Diagram $(a)$ represents a single-nucleon contribution, $(b)$ a momentum-dependent radius correction, and $(c)$ the coupling via a two-body current.}
	\label{fig:diagrams}
\end{figure}

\begin{figure*}[t] 
	\centering
	\includegraphics[width=.9\textwidth,clip]{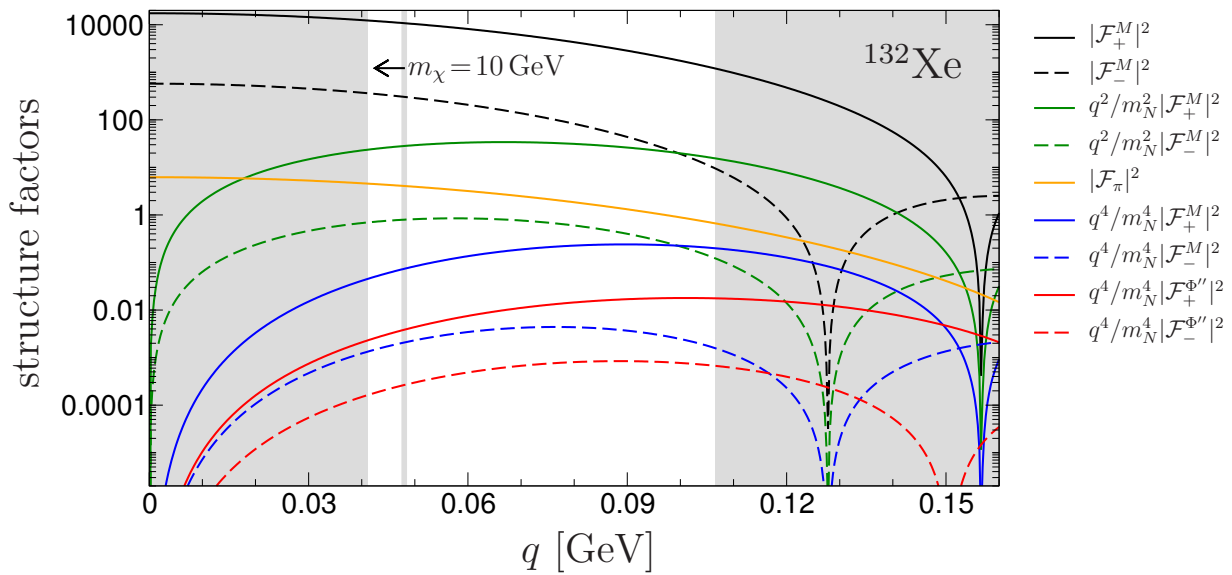}
	\caption{Nuclear structure factors for $^{132}$Xe from~\cite{Hoferichter:2016nvd}, see main text for details. Solid/dashed lines refer to isoscalar/isovector nucleon couplings. The energy thresholds at $6.6\keV$--$43.3\keV$, from the search window of the XENON100 detector, translate into the solid gray bands indicating the momentum transfers below the lower (above the upper) detector threshold at $q_\text{thr}^\text{low}=0.0412\GeV$ ($q_\text{thr}^\text{up}=0.1065\GeV$). The vertical gray line shows the estimated maximum momentum transfer for a WIMP mass $\mc=10\GeV$.
    For $\mc=100\GeV$ and $\mc=1\TeV$ the maximum momentum transfer exceeds $q_\text{thr}^\text{up}$.
    The ordering of the solid lines in the legend follows the ordering of the structure factors in the search window of the detector.}
	\label{fig:structure_factors}
\end{figure*}

Subleading corrections to WIMP--nucleus scattering are most conveniently analyzed in chiral EFT~\cite{Cirigliano:2012pq,Menendez:2012tm,Klos:2013rwa,Baudis:2013bba,Cirigliano:2013zta,Vietze:2014vsa,Hoferichter:2015ipa,Hoferichter:2016nvd,Korber:2017ery,Hoferichter:2017olk}  
(see also related work on WIMP--nuclear response calculations
of $A \leq 4$ nuclei~\cite{Gazda:2016mrp} and on using chiral EFT for WIMP--nucleon interactions~\cite{Bishara:2016hek,Bishara:2017pfq}),
and can be classified into the three categories shown in Fig.~\ref{fig:diagrams}. First, diagram $(a)$ represents single-nucleon contributions as they would appear in NREFT. These corrections have been cataloged in terms of NREFT operators $\Op_i$ in~\cite{Fan:2010gt,Fitzpatrick:2012ix,Anand:2013yka}, leading to a distinct set of nuclear responses, canonically referred to as $M$, $\Delta$, $\Sigma'$, $\Sigma''$, $\tilde \Phi'$, $\Phi''$.
In particular, the isoscalar $M$ response is related to the traditional Helm form factor. Out of the other five responses, only $\Phi''$ can be enhanced by the coherent contribution of a large number of nucleons.
Next, there are so-called radius corrections, which in chiral EFT are related to momentum-dependent loop effects, see diagram $(b)$ in Fig.~\ref{fig:diagrams}. These terms depend on the same set of nuclear structure factors as the precedent single-nucleon contributions, but are suppressed with respect to them by powers of the (small) momentum transfer $q$. Finally, there are contributions due to two-body currents, diagram $(c)$, which represent the coupling of the WIMP to two nucleons, and lead to a genuinely new set of structure factors.

The leading contributions to coherent WIMP--nucleus scattering were analyzed in detail in~\cite{Hoferichter:2016nvd}. When retaining the dominant contributions that are coherently enhanced, the cross section 
can be written as
\begin{align}
\label{structure_factors}
\frac{\diff \sigma}{\diff q^2}&=\frac{1}{4\pi v^2}\bigg|\sum_{I=\pm}\Big(c_I^M-\frac{q^2}{m_N^2} \, \dot c_I^M\Big)\F_I^M(q^2)\notag\\
&\qquad+c_\pi \F_\pi(q^2)+\frac{q^2}{2\mN^2}\sum_{I=\pm}c_I^{\Phi''}\F_I^{\Phi''}(q^2)\bigg|^2\notag\\
&+\frac{1}{4\pi v^2}\bigg|\sum_{I=\pm}\frac{q}{2\mc}\tilde c_I^M\F_I^M(q^2)\bigg|^2.
\end{align}
Here, the structure factors for the $M$ and $\Phi''$ responses, $\F_\pm^M$ and $\F_\pm^{\Phi''}$, describe a coupling to the nucleons that can be isoscalar (same for protons and neutrons, denoted by $+$) or isovector (opposite coupling, represented by $-$). The isoscalar $M$-response is related to the standard Helm form factor, i.e., $\F_\text{SI}$ from Eq.~\eqref{SI_simp} coincides with $\F_+^M$.
$\F_\pi$ originates from the two-body corrections from diagram $(c)$ in Fig.~\ref{fig:diagrams}, which originates from the coupling of the WIMP to the pion. The $q^2$-suppressed terms in the first line of Eq.~\eqref{structure_factors} give the radius corrections, while the remaining terms are produced by subleading NREFT operators. Following~\cite{Hoferichter:2016nvd}, we have kept the two numerically dominant ones, which are $\Op_3$ (leading to the $\Phi''$ response) and $\Op_{11}$ (generating the $M$ response in the last line that does not interfere with the rest). There are two additional coherent NREFT operators, $\Op_{5}$ and $\Op_{8}$, but the functional form of the associated structure factors is very similar to $M$ and to the response corresponding to $\Op_{11}$, respectively. The only difference is an additional dependence on the relative velocity between the WIMP and the nucleus.

Reference~\cite{Hoferichter:2016nvd} provides the structure factors $\F_\pm^M$, $\F_\pm^{\Phi''}$, and $\F_\pi$ used in this work. The results were based on large-scale state-of-the-art calculations for all stable xenon isotopes using the nuclear shell model. The Schr\"odinger equation was explicitly solved in a configuration space consisting of the five single-particle orbitals, for protons and neutrons, with energies closer to the Fermi level. Meanwhile the lowest energy orbitals were considered completely filled by nucleons. As a test, the excitation spectra of xenon isotopes were shown to be in very good agreement with experiment. The one-body structure factors $\F_\pm^M$ and $\F_\pm^{\Phi''}$ were obtained using the full wave functions obtained in the nuclear shell-model calculations, while for $\F_\pi$ only the diagonal contributions were considered, but taking into account the calculated shell-model occupancies.
References~\cite{Vietze:2014vsa,Co:2012adt} showed that the relevant coherent structure factors at low momentum transfer are not very sensitive to the nuclear structure details of the nuclei involved, suggesting that the associated uncertainties may be less relevant than the astrophysical ones~\cite{Cerdeno:2012ix}.

All structure factors are accompanied by coefficients $c$ that subsume hadronic matrix elements and Wilson coefficients describing the interaction of the WIMP with standard model fields. For the isoscalar $M$-response, the exact same coefficient, $c_+^M$, also appears in the WIMP--nucleon cross section $\sigma_0$, which is why the WIMP--nucleus cross section can be brought into the simple form given by Eq.~\eqref{SI_simp}.     

In the event of a detection of WIMPs scattering in a direct detection experiment, the nature of the WIMP--nucleus interaction could in principle
be probed by the distinct functional forms of the different terms in Eq.~\eqref{structure_factors}. 
The $q$-dependence of the subleading corrections can differ from $\F_+^M$ either due to a completely independent nuclear structure factor or due to additional kinematic suppression in $q$. In a scenario where the leading response vanishes, $c_+^M=0$, in principle several different contributions could vie for second place, and in general a global analysis would be required. However, for practical reasons, within a single experiment, such a global approach may not be feasible, so that in a first step we study to which extent each of the subleading corrections can be distinguished from $|\F_+^M|^2$. 
In such a one-operator-at-a-time approach, the decomposition in Eq.~\eqref{structure_factors} shows that the relevant structure factors are $|\F_-^M|^2$, $q^4/\mN^4 |\F_\pm^M|^2$, $|\F_\pi|^2$, $q^4/4\mN^4 |\F_\pm^{\Phi''}|^2$, and $q^2/4\mc^2 |\F_\pm^M|^2$.
The coefficients $c$ will be adjusted in such a way that the overall WIMP scattering rate is consistent with the published SI limits on $\sigma_0$,
so that the numerical factors do not play a role.
The results from~\cite{Hoferichter:2016nvd} for the various structure factors are shown in Fig.~\ref{fig:structure_factors}. Since only the functional form of each structure factor is relevant for our study, in Fig.~\ref{fig:structure_factors} the numerical factors have been dropped and every power of $q$ is accompanied simply by $\mN$.

\section{Experimental assumptions and method}
\label{sec:method}

Once a signal excess is found by one of the leading direct dark matter detection experiments, the immediate question to be answered concerns the nature of the WIMP--nucleus interaction. While a standard interpretation would assume an isoscalar SI interaction described phenomenologically by a Helm form factor, several other interactions are also possible, as discussed in Sec.~\ref{sec:theory}. The underlying interaction could in principle be distinguished from the standard Helm assumption due to the distinct $q$-dependence of each structure factor. In practice, the feasibility of this discrimination depends critically on how many signal events are observed. Besides the exposure, the number of events detected in an experiment is a function of the strength of the WIMP--nucleus interaction (which can be expressed in a cross-section equivalent to the standard SI WIMP--nucleon interaction $\sigma_0$), the WIMP mass $\mc$, and the properties of the detector, mainly its energy thresholds, background rate and distribution.

\subsection{Detector assumptions}

In order to not rely on detailed properties of different detectors, we assume a XENON100-like detector (details in~\cite{Aprile:2012vw,Aprile:2011dd}), with the settings based on the data from 2011 to 2012~\cite{Aprile:2012nq} throughout the whole analysis. The choice is motivated by the fact that there is not enough information publicly available to reconstruct the signal and background model from XENON1T, while for XENON100 dedicated published analyses provide the needed insight, e.g., see Ref.~\cite{Aprile:2017aas}.
The XENON100 settings refer to any experimental properties such as the background composition and the given lower energy threshold, as well as the chosen search window causing the upper energy threshold. On the contrary, we scale linearly the total active mass of the experiment $M$ in order to extend our analysis to state-of-the-art experiments such as XENON1T and future ones like XENONnT and the proposed DARWIN. We also adjust the absolute background rate to each experiment, as discussed below.

The detectors considered in our analysis use a dual-phase TPC to observe the nuclear recoils (NR) that result from the WIMP scattering off a xenon nucleus. A dual-phase TPC is sensitive to two different signals: direct scintillation light (S1) and delayed electroluminescence photons caused by initial ionization (S2). The ratio of the strength of the two signals -- measured in terms of the total deposited energy, in units of the number of produced photoelectrons (PE) -- depends on whether the particle interacts directly with the nucleus, as expected for WIMPs, or with the electrons within the atomic shell producing electronic recoils (ER), as expected for the dominating background such as $\gamma$-rays or $\beta$-particles. The S1 and S2 signals are anti-correlated due to the nature of the production mechanism \cite{Aprile:2009dv}. Several effects need to be included in order to reconstruct the true deposited energy, and the corrected signals are usually referred to as cS1 and cS2. These corrections include various detector effects (e.g., the geometrical position of the interaction) that need to be incorporated using calibration sources. Furthermore it has to be taken into account that the value of the created number of PE per unit energy is not constant but follows a Poissonian distribution and thus, the $q$-dependence of any signal distribution will be smeared out when observed experimentally.

Another important experimental property of a detector is its energy threshold, as the lowest-energy NRs usually escape detection. As the differential NR spectrum corresponding to a standard isoscalar SI 
WIMP--nucleus interaction is a falling exponential, the highest NR rate is expected at the lowest energies~\cite{Undagoitia:2015gya}. Accordingly, for a Helm form factor a significant fraction of information is lost below the detector's energy threshold, especially compared with a structure factor that vanishes at $q=0$. Nonetheless, as discussed in the previous paragraph, in practice the detector threshold is not a fixed line in energy space but it is smeared out. Figure~\ref{fig:structure_factors} shows indicative momentum-transfer thresholds for the experiments studied in this work.

In order to account for the different experimental settings the absolute background rate $b$ of each experiment is obtained. Therefore, we determine the background scaling from the ratio of ER background rates corresponding to XENON100 and the respective experiment, considering the search region of XENON100. As the rate for XENON100 has been determined to be $5.3\times 10^{-3} \frac{\text{evts}}{\text{kg} \times \text{keV} \times \text{day}}$~\cite{Aprile:2012nq} and the rate for XENON1T has been measured as $1.9\times 10^{-4} \frac{\text{evts}}{\text{kg} \times \text{keV} \times \text{day}}$~\cite{Aprile:2017iyp}, we take for XENON1T $b_\text{XENON1T}=0.036\,b_\text{XENON100}$. For XENONnT the background expectations given in \cite{Aprile:2015uzo} are used with the exception of the expected further background reduction by a factor of 100 for the dominant backgrounds induced by $^{222}$Rn daughters. As this is too optimistic given the experience from XENON1T, we use instead a reduction factor of 10 with respect to XENON1T, leading to $b_\text{XENONnT}=0.0036\,b_\text{XENON100}$. For DARWIN we have a rate of $5.8\times 10^{-6} \frac{\text{evts}}{\text{kg} \times \text{keV} \times \text{day}}$  taken from~\cite{Aalbers:2016jon}, which leads to a background rate $b_\text{DARWIN}=0.0011\,b_\text{XENON100}$. Here we only consider the ER rate from xenon-intrinsic contamination, as this is the dominant background for the cross sections considered in our study.

Our choice of background neglects a few aspects which, if taken into account, would increase the sensitivity of our study. First, the rate of NR backgrounds, which is the challenging component when it comes to the differentiation of two signal models, decreases more pronouncedly than the ER background. Second, the composition of the NR background changes when moving from XENON100 to current or future experiments. For instance, for XENON1T it is already known that neutrons from muon-induced scatters do not contribute anymore to the overall background spectrum due to the active muon veto~\cite{Aprile:2014zvw}. Another example is that the contribution of radiogenic neutrons decreases with increasing detector volume, and additionally future experiments actually consider reducing this background source by active neutron veto techniques \cite{Aalbers:2016jon}. Finally, at higher sensitivities coherent neutrino--nucleus scattering
occurs, further changing the composition of the background in a way that depends on the source of the neutrinos and their respective energy. Besides these differences, there are also experimental properties that are different among the XENON100-like detector we consider for reference and future ones. One example is the light yield -- the number of detected PE per unit energy -- which increases by more than a factor of 2 from XENON100 to XENON1T. Overall, all these effects combined suggest that the results obtained in our analysis may underestimate the capability of present and especially future experiments to discriminate between different WIMP--nucleon interactions.

\subsection{Monte Carlo and likelihood analysis}

\begin{figure}
	\centering
	\includegraphics[width=\columnwidth]{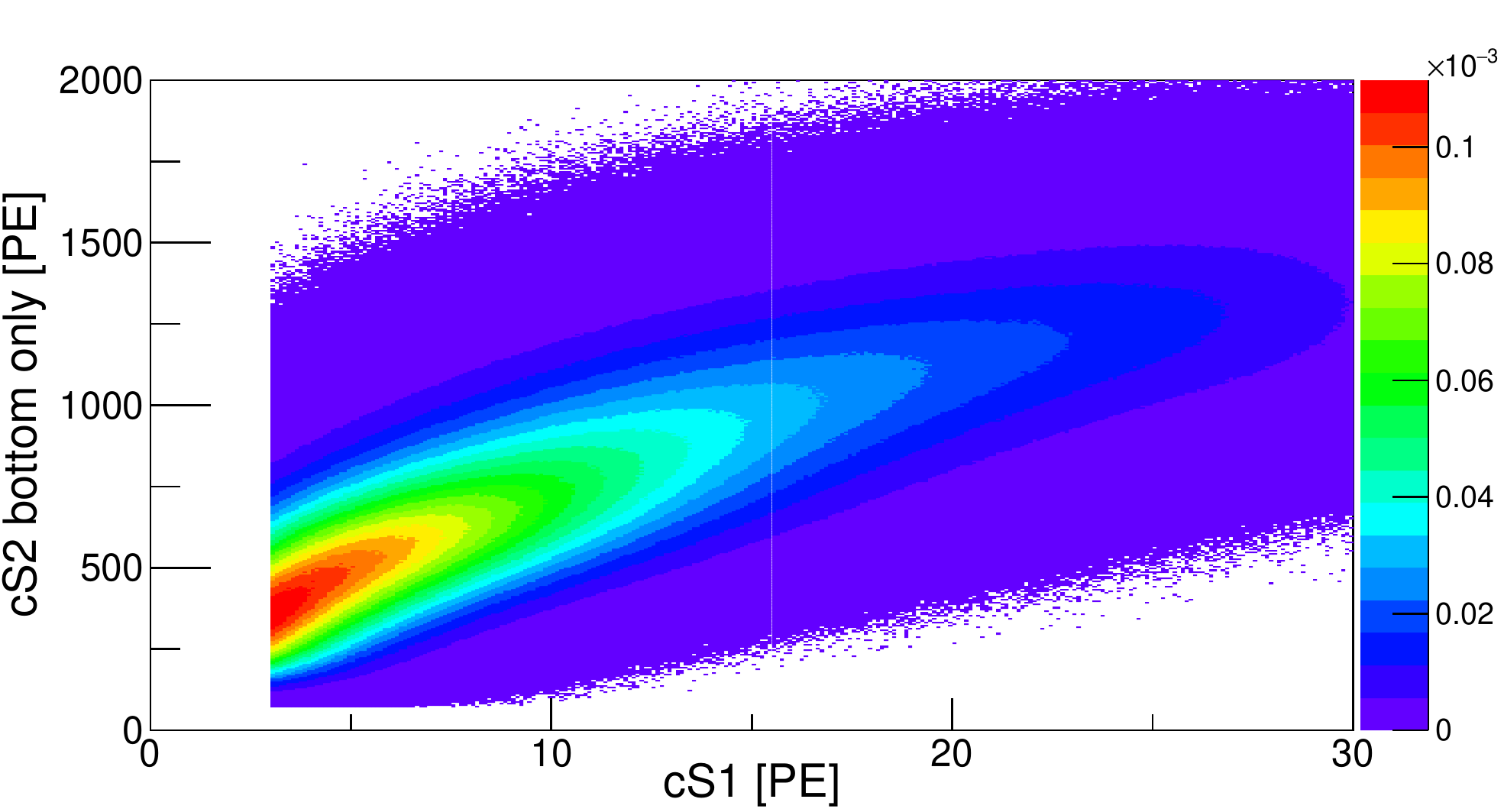}
	\includegraphics[width=\columnwidth]{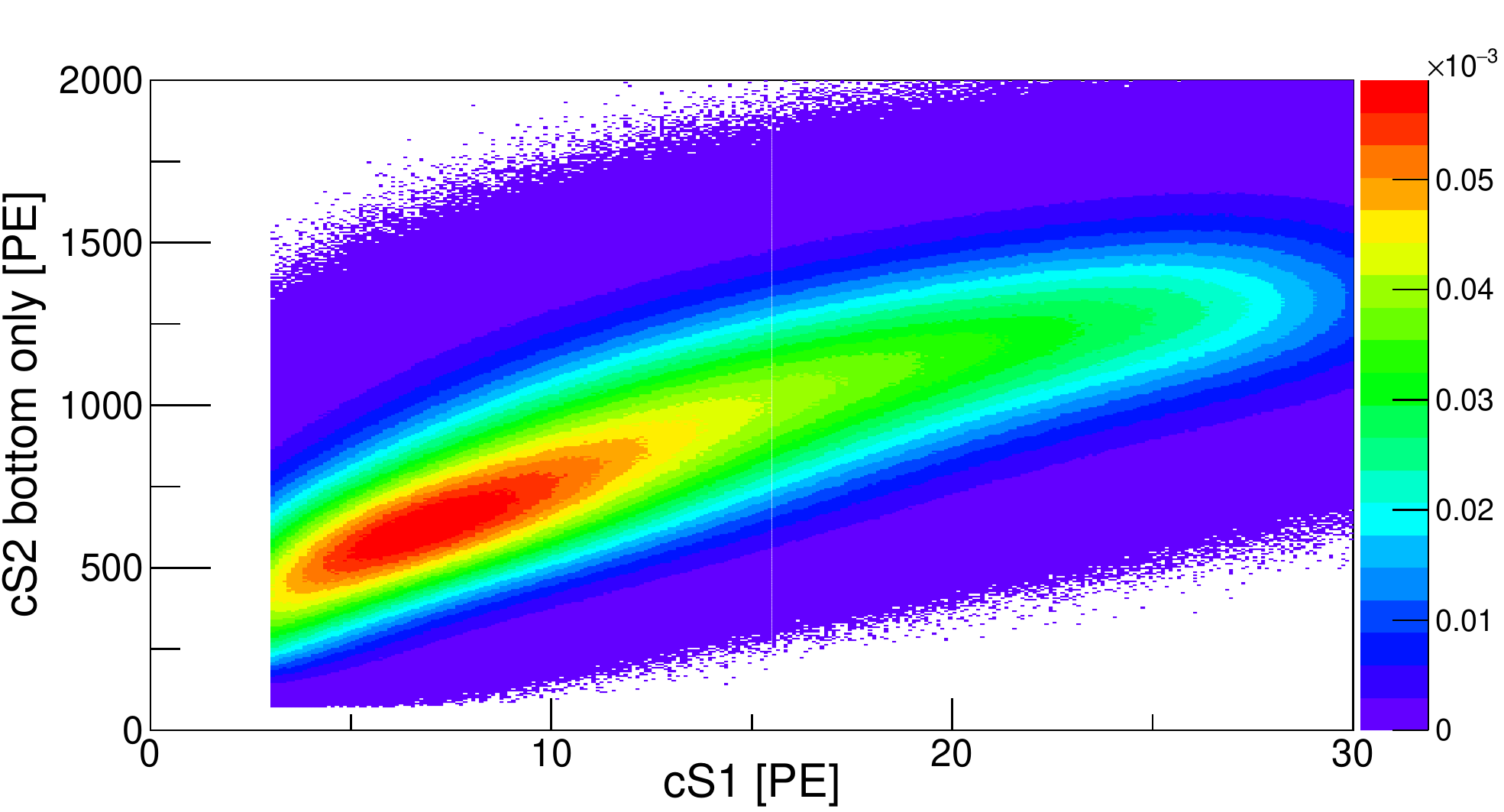}
	\caption{Top: Probability density function based on the signal model using the standard isoscalar structure factor $|\mathcal{F}^M_+|^2$, corresponding to the Helm form factor. The parameter space shows the scintillation signal (cS1) vs.\ the part of the ionization signal (cS2) that is seen in the bottom of the detector. The color code denotes a probability for each bin and the function is normalized to 1 in the given parameter space. Bottom: Probability density function for the signal with the underlying structure factor $q^2/4\mc^2 |\F_+^M|^2$ as well normalized to 1 in the given parameter space.}
	\label{fig:sig_mod_q_diff}
\end{figure}

We model with a toy-MC approach
the different signals corresponding to the underlying nuclear structure factors discussed in Sec.~\ref{sec:theory}. In our analysis we fix $\mc$ and chose a specific $\sigma_0$, based on the Helm form factor, that corresponds to a certain interaction strength and leads to a number of expected WIMP events $\langle N_\text{sig}\rangle$. $\langle N_\text{sig}\rangle$ is a function of exposure $M \times t$ (with $t$ the time), and can be calculated given the energy window of 3\,PE\,to\,30\,PE in the scintillation channel (cS1), corresponding to $6.6\keV$--$43.3\keV$ for NR (see Fig.~\ref{fig:structure_factors}). So while we keep $\langle N_\text{sig}\rangle$ constant when investigating the different underlying nuclear structure factors one-at-a-time, the shape of the signal model will differ in the detector signal parameter space (the cS1-cS2-plane) due to the different $q$-dependence of each term. An example of these differences is shown in Fig.~\ref{fig:sig_mod_q_diff}. 

Likewise, the background rate ($b_\text{XENON100}$, $b_\text{XENON1T}$, $b_\text{XENONnT}$, or $b_\text{DARWIN}$) determines the expected number of background events $\langle N_\text{bkg}\rangle$, as a function of exposure $M \times t$. Using the mean number of signal and background events,  we generate experimental outcomes via toy-MC simulations. The resulting distribution of events from the toy-MC is then compared to probability density functions (pdfs) derived from the background model of XENON100 in combination with a) the signal model assuming the Helm form factor ($f_\text{Hs+b}$), and b) the signal model assuming the non-Helm form factor of interest ($f_\text{nHs+b}$). These total pdfs are generated by weighting the chosen signal model ($f_\text{Hs}$ or $f_\text{nHs}$) and the background model ($f_\text{b}$) with their expected mean number of events and normalized to the total number of events as described by
\beq
f_\text{Hs+b/nHs+b} = \frac{\langle N_\text{sig}\rangle \times f_\text{Hs/nHs} + \langle N_\text{bkg}\rangle \times f_\text{b}}{\langle N_\text{sig}\rangle+\langle N_\text{bkg}\rangle} \,.
\label{eq:combined_pdfs}
\eeq

In order to decide whether a structure factor can be distinguished
from the Helm form factor we perform an unbinned likelihood-ratio test, where the hypothesis of a Helm form factor signal acts as the null model and the hypothesis of the respective different structure factor as the alternative model. Likewise we perform an analysis in an analogous manner to check that the non-standard structure factor can also be distinguished from pure background.

\begin{figure}
	\centering
	\includegraphics[width=\columnwidth]{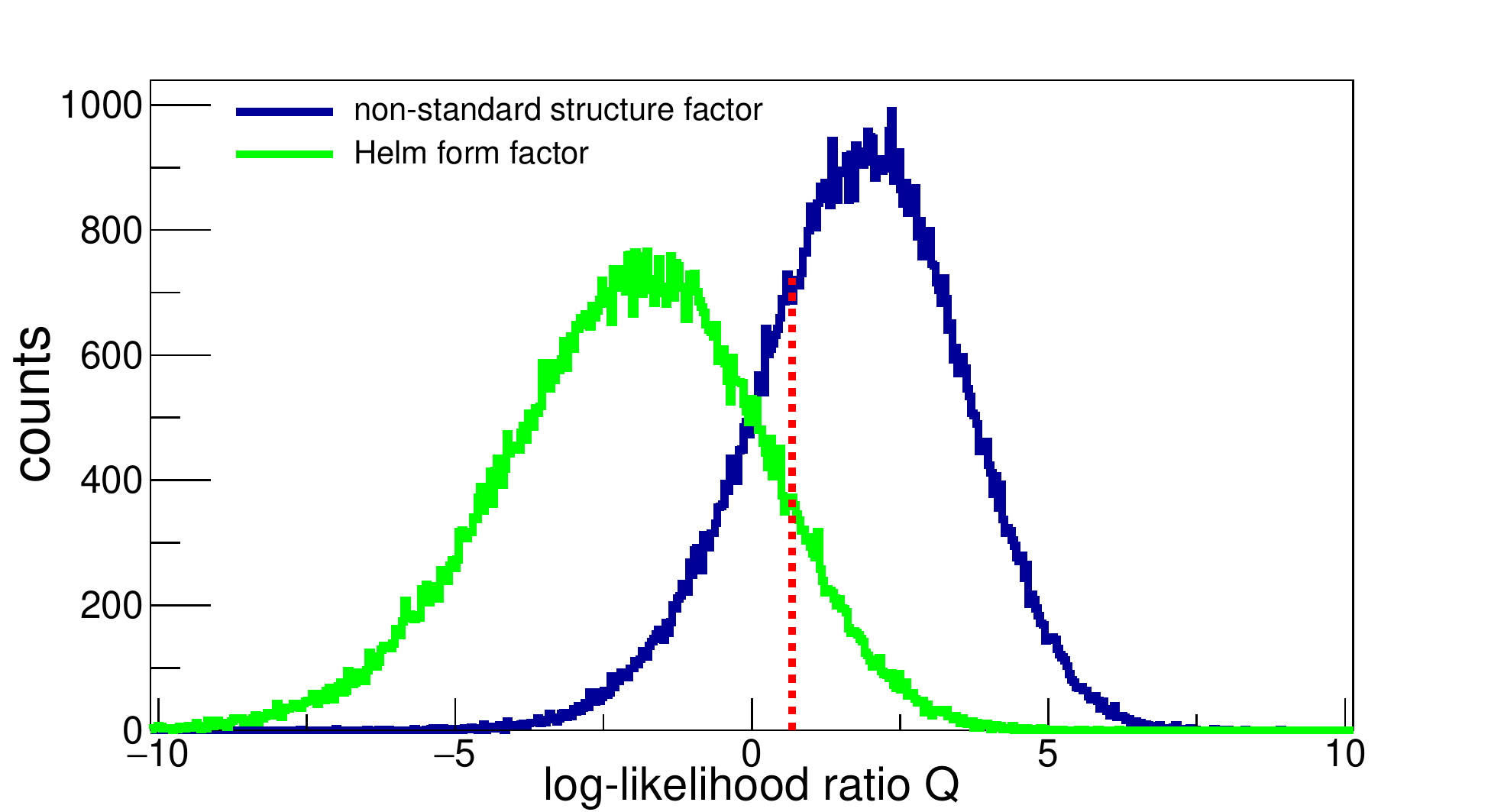}
	\caption{Log-likelihood ratio distributions for the null distribution (green, based on the Helm form factor) and the alternative distribution (blue, based on a non-standard structure factor). Marked in red is the $1.28\,\sigma$ line for the null distribution. The discriminating power is given by integrating the alternative distribution from the reference red line towards infinity. In this case this yields a discrimination power of $74\%$.}
	\label{fig:example_distribution}
\end{figure}

In detail, the procedure is as follows. For a total number of events $N_\text{tot} = \langle N_\text{sig}\rangle + \langle N_\text{bkg}\rangle$ we first determine the null distribution by running toy-MC experiments generating events $x_i$ from the Helm form factor signal with background, and compute the total likelihood-ratio $Q_{N_\text{Hs}}$
\beq
Q_{N_\text{Hs}} = \prod_{i=1}^{N_\text{tot}}\frac{p(x_i|f_\text{nHs+b})}{p(x_i|f_\text{Hs+b})} \,,
\eeq
where $p$ is the probability of the event given the respective probability density function $f$. By generating this ratio for each of the $n$-trial in the toy-MC we obtain a null distribution, shown by the green curve in Fig.~\ref{fig:example_distribution}, where we determine the $1.28\,\sigma$ line corresponding to the one-sided 90$\%$ quantile. 

Next we create a data set with events $y_i$, distributed according to the non-Helm form factor signal model of interest and the same background model as before. We obtain the likelihood-ratio by
\beq
Q_{{N_\text{nHs}}} = \prod_{i=1}^{N_\text{tot}}\frac{p(y_i|f_\text{nHs+b})}{p(y_i|f_\text{Hs+b})}\,.
\eeq
For each trial $n$ we compare the same hypothesis as in the previous step. This way we generate an alternative distribution with $n$ entries, shown as the blue curve in Fig.~\ref{fig:example_distribution}, to be compared to the previous distribution based on the standard Helm form factor. By determining the fraction of entries above the previously-derived $1.28\,\sigma$ line, we can make a statement about the discrimination power of present and future experiments for the particular non-standard structure factor studied. 

Finally, we follow the same  procedure to test each non-standard structure factor signal against the null hypothesis of ``background only.'' This ensures that the corresponding signal would be distinguishable from background.

\section{Results}
\label{sec:results}

We study three different WIMP masses: $\mc=10\GeV$, $\mc=100\GeV$, and $\mc=1\TeV$. However, as seen in Fig.~\ref{fig:structure_factors} for the lightest $\mc=10\GeV$ there is only a narrow corridor of allowed momentum transfers. As a result, for this WIMP mass we could not obtain any discriminating power for any of the structure factors in Fig.~\ref{fig:structure_factors}, even in the most optimistic scenario of the future DARWIN experiment. For the other two WIMP masses\footnote{In principle direct-detection dark matter experiments cannot accurately reconstruct the WIMP mass from the data for large WIMP masses. However, the features in the recoil spectra, which allow one to differentiate between the different structure factors, do not depend much on the WIMP mass. Tables~\ref{tab:XE1T}--\ref{tab:Darwin} illustrate that even the large differences in WIMP masses of $\mc=100\GeV$ and $\mc=1\TeV$ do not change the discrimination power significantly. Therefore, we did not scan in detail over the entire WIMP mass range but chose exemplary WIMP masses in our study.} 
we considered interaction strengths based on different values of the SI cross section, starting from the reach of today's sensitivities and moving towards lower cross sections as long as a discrimination is possible.
We took the starting values from the limit in \cite{Aprile:2017iyp}, corresponding to $\sigma_0= 10^{-46}\,\text{cm}^2$ for $\mc=100\GeV$ and $\sigma_0=10^{-45}\,\text{cm}^2$ for $\mc=1\TeV$. We checked the discrimination power as a function of exposure for a XENON1T-like experiment up to 10\,ton\,years, for a XENONnT-like (also LZ-like or PandaX-xT-like) experiment  up to 30\,ton\,years, and for a DARWIN-like experiment up to 200\,ton\,years. 

\begin{figure}[t] 
	\centering
	\includegraphics[width=\columnwidth]{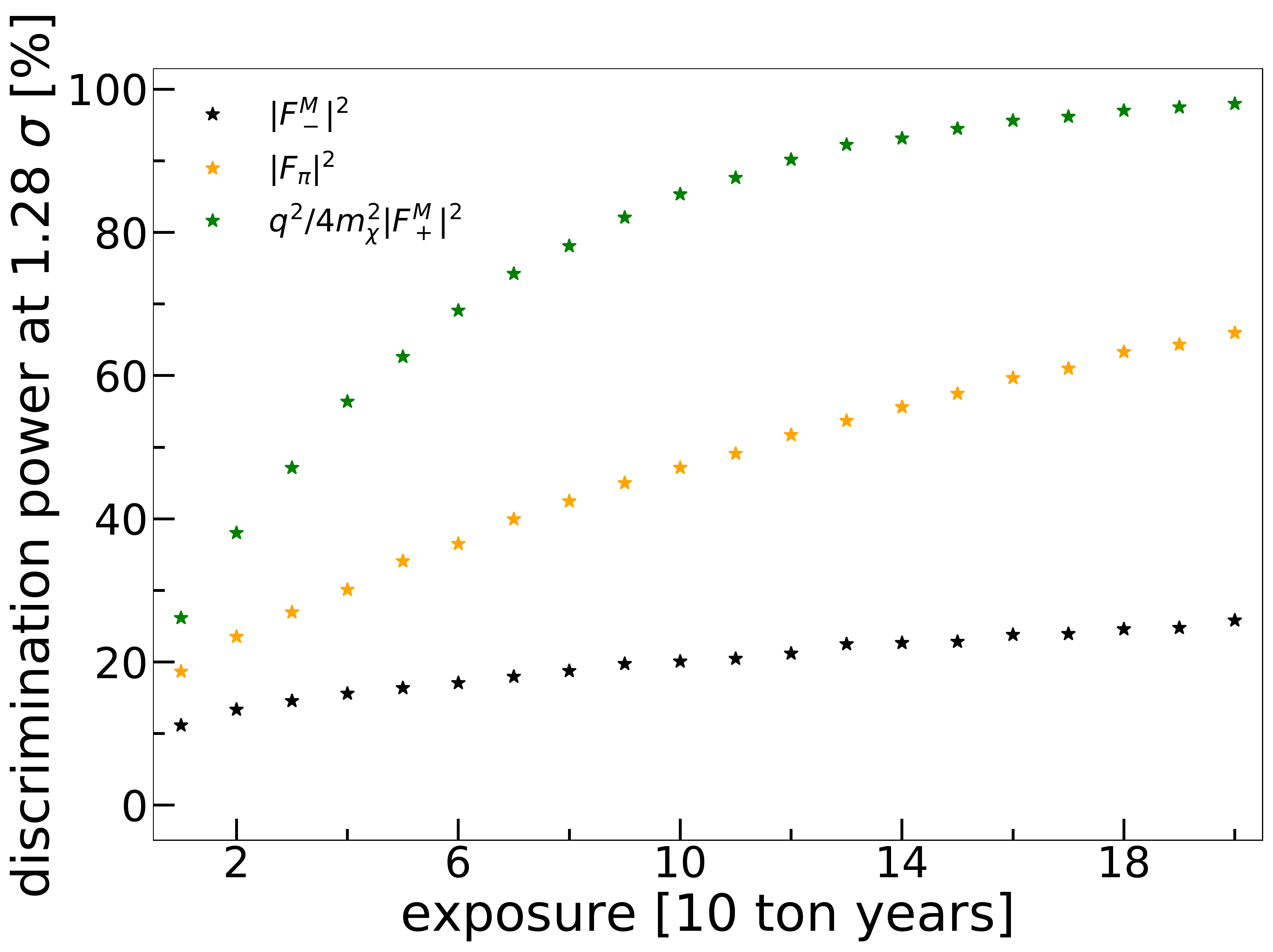}
	\caption{Discrimination power vs.\ exposure for three selected structure factors, $|\F_-^M|^2$ (black), $|\F_\pi|^2$ (orange), and $q^2/4\mc^2 |\F_-^M|^2$ (green). The detector setting is DARWIN-like, with $\mc=100\GeV$ and $\sigma_0=10^{-47}\,\text{cm}^2$.} 
	\label{fig:results_example}
\end{figure}

\begin{table}[t]
	\centering
	\renewcommand{\arraystretch}{1.3}
	\setlength{\tabcolsep}{0.12cm}
	\caption{Discrimination power (in $\%$) of the XENON1T settings after 10 ton years of exposure.}
	\label{tab:XE1T}
	\begin{tabular}{c|cc|cc}
		\toprule
		$\mc$ & \multicolumn{2}{c}{$100\GeV$} & \multicolumn{2}{c}{$1\TeV$}\\
		$\sigma_0\,[\text{cm}^2]$ 
		& $10^{-46}$ & $10^{-47}$ & $10^{-45}$ & $10^{-46}$ \\
		\colrule
		$|\F_-^M|^2$ & $17$ & $10$ & $21$ & $11$\\ 
		$q^2/4\mc^2 |\F_+^M|^2$ & $94$ & $19$ & $98$ & $20$ \\
		$q^2/4\mc^2 |\F_-^M|^2$ & $74$ & $16$ & $90$ & $17$ \\
		$q^4/\mN^4 |\F_+^M|^2$ & $100$ & $25$ & $100$ & $28$ \\ 
		$q^4/\mN^4 |\F_-^M|^2$ & $100$ & $23$ & $100$ & $25$ \\
		$|\F_\pi|^2$ & $38$ & $13$ & $48$ & $14$\\
		$q^4/4\mN^4 |\F_+^{\Phi''}|^2$ & $100$ & $26$ & $100$ & $30$ \\
		$q^4/4\mN^4 |\F_-^{\Phi''}|^2$ & $100$ & $25$ & $100$ & $29$ \\
		\botrule
	\end{tabular}
	\renewcommand{\arraystretch}{1.0}
\end{table}

\begin{table}[t]
	\centering
	\renewcommand{\arraystretch}{1.3}
	\setlength{\tabcolsep}{0.12cm}
	\caption{Discrimination power (in $\%$) of a XENONnT-like experiment after 30 ton years of exposure.}
	\label{tab:XEnT}
	\begin{tabular}{c|ccc|ccc}
		\toprule
		$\mc$ & \multicolumn{3}{c}{$100\GeV$} & \multicolumn{3}{c}{$1\TeV$}\\
		$\sigma_0\,[\text{cm}^2]$ 
		& $10^{-46}$ & $10^{-47}$ & $10^{-48}$ & $10^{-45}$ & $10^{-46}$ & $10^{-47}$\\
		\colrule
		$|\F_-^M|^2$ & $37$ & $13$ & $10$ & $21$ & $15$ & $11$\\ 
		$q^2/4\mc^2 |\F_+^M|^2$ & $100$ & $59$ & $14$ & $100$ & $71$ & $15$\\
		$q^2/4\mc^2 |\F_-^M|^2$ & $100$ & $39$ & $13$ & $100$ & $53$ & $13$\\
		$q^4/\mN^4 |\F_+^M|^2$ & $100$ & $90$ & $16$ & $100$ & $95$ & $17$\\ 
		$q^4/\mN^4 |\F_-^M|^2$ & $100$ & $81$ & $15$ & $100$ & $87$ & $16$\\
		$|\F_\pi|^2$ & $89$ & $23$ & $12$ & $98$ & $28$ & $12$\\
		$q^4/4\mN^4 |\F_+^{\Phi''}|^2$ & $100$ & $93$ & $17$ & $100$ & $98$ & $19$\\
		$q^4/4\mN^4 |\F_-^{\Phi''}|^2$ & $100$ & $89$ & $17$ & $100$ & $96$ & $17$\\
		\botrule
	\end{tabular}
	\renewcommand{\arraystretch}{1.0}
\end{table}

\begin{table}[t]
	\centering
	\renewcommand{\arraystretch}{1.3}
	\setlength{\tabcolsep}{0.12cm}
	\caption{Discrimination power (in $\%$) of a DARWIN-like experiment after 200 ton years of exposure.}
	\label{tab:Darwin}
	\begin{tabular}{c|ccc|ccc}
		\toprule
		$\mc$ & \multicolumn{3}{c}{$100\GeV$} & \multicolumn{3}{c}{$1\TeV$}\\
		$\sigma_0\,[\text{cm}^2]$
		& $10^{-46}$ & $10^{-47}$ & $10^{-48}$ & $10^{-45}$ & $10^{-46}$ & $10^{-47}$\\
		\colrule
		$|\F_-^M|^2$ & $94$ & $26$ & $12$ & $100$ & $35$ & $13$\\ 
		$q^2/4\mc^2 |\F_+^M|^2$ & $100$ & $100$ & $34$ & $100$ & $100$ & $41$\\
		$q^2/4\mc^2 |\F_-^M|^2$ & $100$ & $98$ & $25$ & $100$ & $100$ & $32$\\
		$q^4/\mN^4 |\F_+^M|^2$ & $100$ & $100$ & $55$ & $100$ & $100$ & $63$\\ 
		$q^4/\mN^4 |\F_-^M|^2$ & $100$ & $100$ & $47$ & $100$ & $100$ & $53$\\
		$|\F_\pi|^2$ & $100$ & $66$ & $17$ & $100$ & $81$ & $20$\\
		$q^4/4\mN^4 |\F_+^{\Phi''}|^2$ & $100$ & $100$ & $58$ & $100$ & $100$ & $69$\\
		$q^4/4\mN^4 |\F_-^{\Phi''}|^2$ & $100$ & $100$ & $55$ & $100$ & $100$ & $64$\\
		\botrule
	\end{tabular}
	\renewcommand{\arraystretch}{1.0}
\end{table}

As an example, Fig.~\ref{fig:results_example} shows the discrimination power as a function of exposure for three non-standard structure factors in the most optimistic DARWIN-like setting, for $\mc=100\GeV$ and $\sigma_0=10^{-47}\,\text{cm}^2$.
$|\F_-^M|^2$ (black) is the least distinguishable case, with only $26\%$ after the full exposure, due to its similar shape as a function of $q$ compared to $|\F_+^M|^2$, see Fig.~\ref{fig:structure_factors}. 
Separating the leading isoscalar and isovector interaction thus likely requires the study of different target nuclei simultaneously, probing automatically different proton-to-neutron ratios, which can vary from $r\approx0.6$ in xenon to $r=1.0$ for a silicon target.    
For $|\F_\pi|^2$ (orange) the discrimination power is larger, reaching up to $66\%$ for the full exposure of 200 ton years. The most promising is to discriminate the structure factor $q^2/4\mc^2 |\F_-^M|^2$ (green), which can be distinguished in $98\%$ of the cases at full exposure. Likewise, all structure functions with a similar momentum dependence show this high discrimination power as seen in Tables~\ref{tab:XE1T}--\ref{tab:Darwin}.
This pattern is fully consistent with the expectation from Fig.~\ref{fig:structure_factors}, i.e., the more the $q$-behavior of a given structure factor differs from $|\F_+^M|^2$, the better the two responses can be discriminated. For that reason, structure factors that vanish at $q=0$ always offer the highest discrimination powers for a given exposure.

Tables~\ref{tab:XE1T}--\ref{tab:Darwin} summarize the results for each of the experimental settings. For WIMP masses $\mc=100\GeV$ and $\mc=1\TeV$ the results give the fraction of the experiments above the $1.28\,\sigma$ line of the null distribution for each of the cross sections $\sigma_0$ considered.
Table~\ref{tab:XE1T} shows that if in the next run XENON1T observes WIMP scattering corresponding to a $\sigma_0$ slightly higher than present limits -- close to $10^{-46}\,\text{cm}^2$ for $\mc=100\GeV$ and $10^{-45}\,\text{cm}^2$ for $\mc=1\TeV$ -- there are good chances of discrimination if the underlying structure factor actually corresponds to a signal that vanishes at $q=0$. 
Table~\ref{tab:XEnT} shows that, in that case, a XENONnT-like experiment would be able to discriminate an $|\F_\pi|^2$ signal as well. A complete discrimination including the isovector $|\F_-^M|^2$ would only be reliable in a DARWIN-like experiment.

\begin{figure}
	\centering
	\includegraphics[width=\columnwidth]{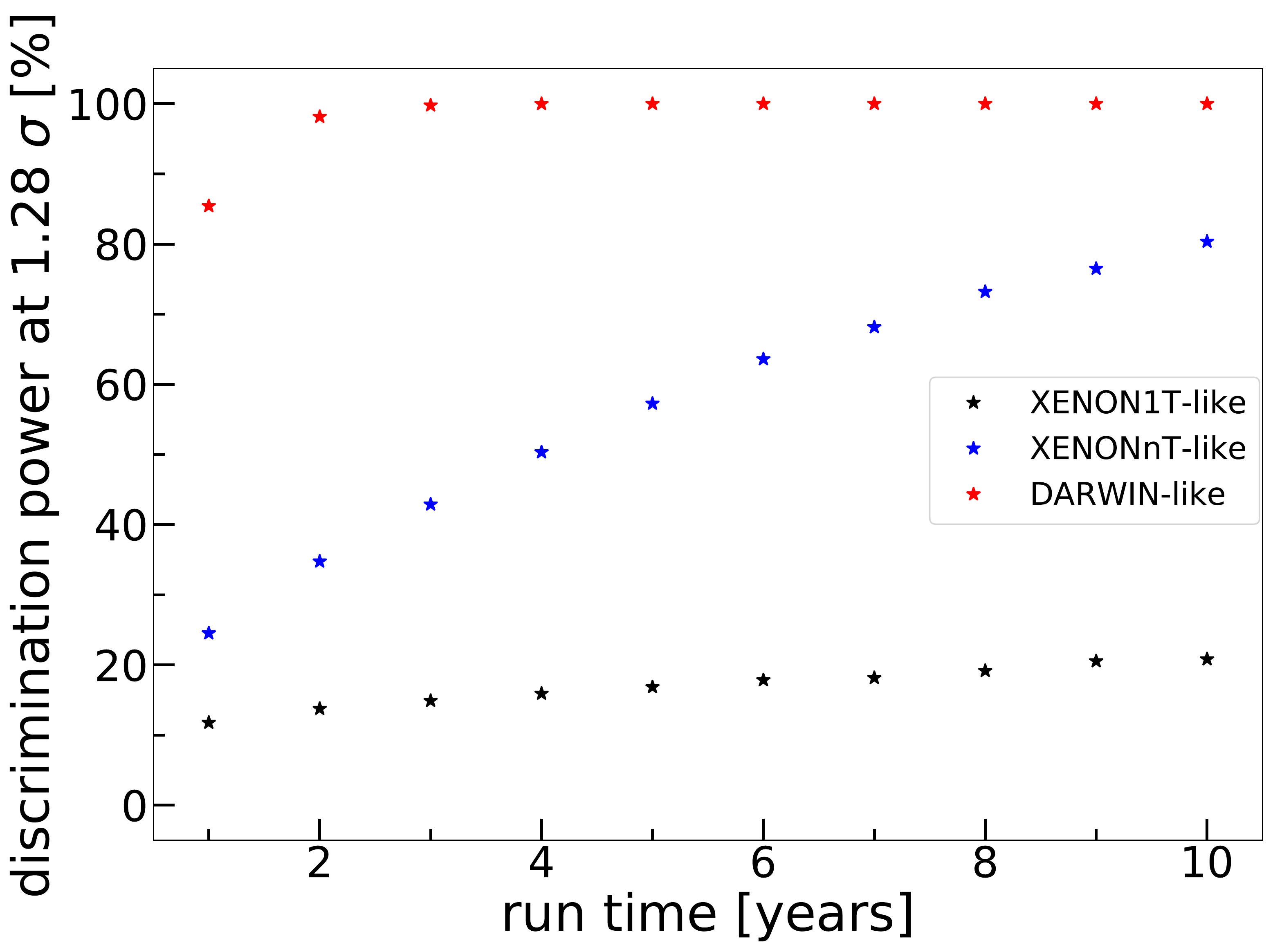}
	\caption{Comparison of the discrimination power during the run time of the three different experimental settings. For XENON1T-like (black) a fiducial target of 1~ton is assumed, for the XENONnT-like setting (blue) a 4~ton fiducial target, and for the DARWIN-like setting (red) we assumed 30~ton fiducial volume. The signal is distributed using the structure factor $q^2/4\mc^2 |\F_+^M|^2$, a WIMP mass $\mc=1\TeV$, and the mean number of events equivalent to $\sigma_0 = 10^{-46}\,\text{cm}^2$.}
	\label{fig:experiment_comparison}
\end{figure}

For $\sigma_0$ values at the reach or slightly beyond XENON1T projected sensitivity -- around $10^{-47}\,\text{cm}^2$ for $\mc=100\GeV$ and $10^{-46}\,\text{cm}^2$ for $\mc=1\TeV$ -- only the future XENONnT would be in a position to discriminate between a Helm form factor and those structure factors suppressed by powers of $q$. A fully reliable discrimination, however, would demand a DARWIN-like experiment, which in addition would also be able to discriminate an underlying $|\F_\pi|^2$ structure factor. This most sensitive detector retains some discrimination power even when $\sigma_0$ is lowered by an additional order of magnitude. The advantage of a DARWIN-like experiment is shown in Fig.~\ref{fig:experiment_comparison}, which illustrates the discrimination power as a function of run time for all three experiment generations in the case of a structure factor $q^2/4\mc^2 |\F_+^M|^2$. 

\begin{figure}[t] 
	\centering
	\includegraphics[width=\columnwidth]{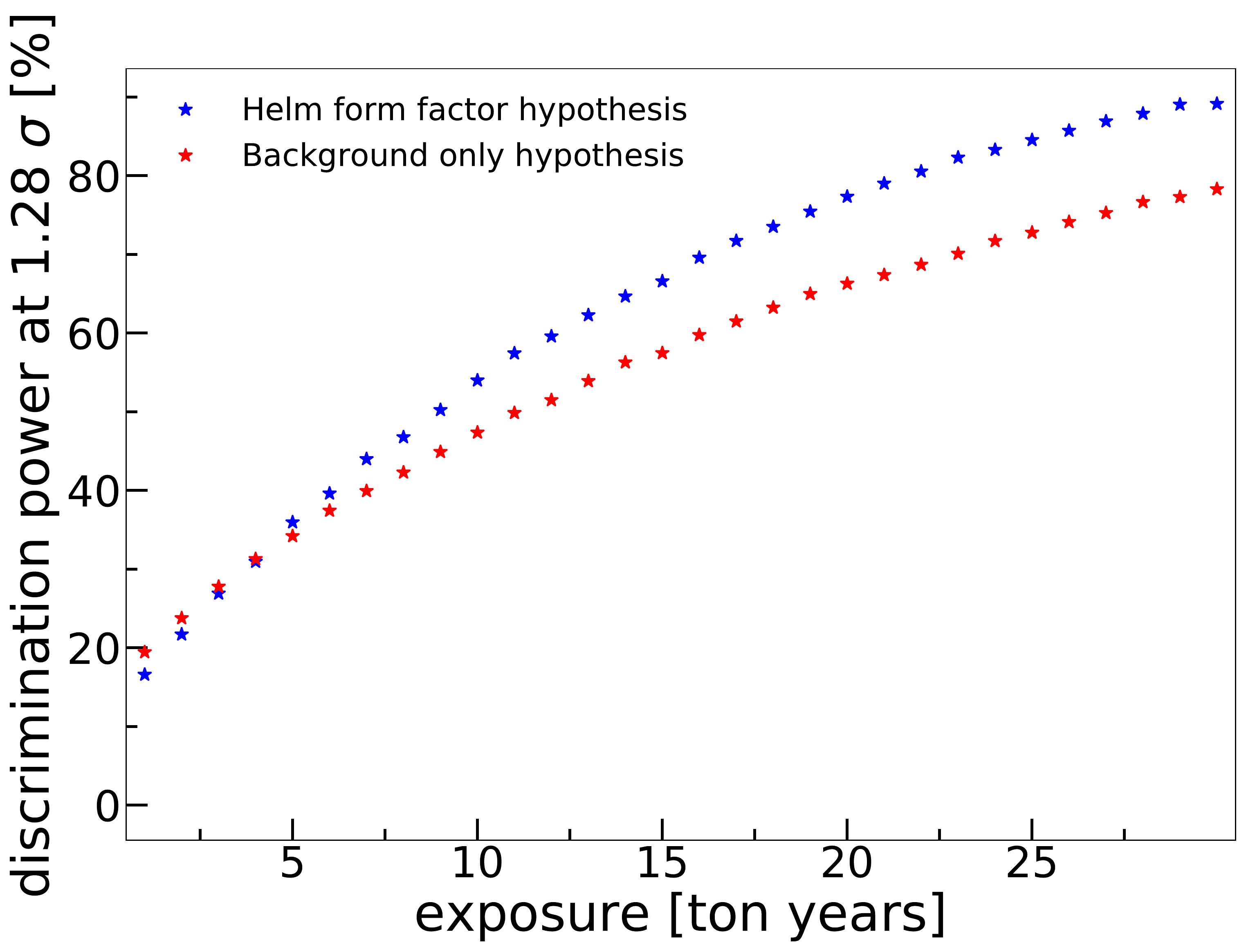}
	\caption{Comparison of the discrimination power of XENONnT for a signal based on the structure factor $q^4/4\mN^4 |\F_-^{\Phi''}|^2$  vs.\ the Helm form factor signal (blue) and for the same signal vs.\ a background-only hypothesis (red). Here $\sigma_0=10^{-47}\text{cm}^2$ and $\mc=100\GeV$.}
	\label{fig:back_example}
\end{figure}

It has to be noted that we show the results for the discrimination power with respect to the Helm form factor in Tables~\ref{tab:XE1T}--\ref{tab:Darwin} regardless of the discrimination power with respect to the pure background. 
This discrimination power of the non-Helm form factors in comparison to the pure background signal hypothesis is larger for all structure factors, with the exception of the nuclear structure factors involving a kinematic $q^4$ suppression. There the capability of discrimination is greater for the Helm form factor signal with background than for the background only scenario, see Fig.~\ref{fig:back_example}.
The observation of this behavior for the $q^4$-dependent structure factors has important consequences for the interpretation of direct-detection limits, since a true dark matter signal could be rejected if its interaction mechanism were based on one of these underlying structure factors and if the Helm form factor signal were the only hypothesis being tested in the experimental analysis. This illustrates that the consideration of subleading responses is not only important to potentially extract further information on the nature of the WIMP, but that even in some cases a signal could be missed if the analysis were restricted to the standard Helm form factor.

\section{Conclusions}
\label{sec:conclusions}

In this paper we have studied the capability of current and future liquid xenon, dark-matter direct-detection detector generations to discriminate between the standard isoscalar SI interaction usually described by a Helm form factor and other coherent, subleading nuclear structure factors.
Starting from a background model for XENON100, we have scaled it to capture the prevalent specifications of XENON1T, XENONnT, and DARWIN detectors. 
Throughout, the generic term ``XENONnT-like'' represents also other dual-phase xenon experiments with a similar sensitivity such as LZ~\cite{Akerib:2015cja} or PandaX-xT~\cite{PandaXxT}.
We have performed toy-MC simulations for dark-matter--nucleus interactions corresponding to
the leading SI interaction as well as for the whole range of most important subleading responses predicted by chiral EFT~\cite{Hoferichter:2016nvd}. Based on these simulations 
we have performed a likelihood analysis that has allowed us to define a metric quantifying the discrimination power of each experimental setting, given a particular exposure and WIMP mass, between the Helm form factor and each of the alternative nuclear structure factors.

In general, this discrimination power is strongly correlated with the difference in shape of the structure factors as a function of the momentum transfer $q$, as well as with the window in $q$ that is effectively probed by a typical detector. Accordingly, for a $\mc=10\GeV$ WIMP the range in $q$ is simply too narrow to achieve any discrimination. However, for heavier WIMPs, considering $\mc=100\GeV$ and $\mc=1\TeV$ as benchmarks, the $q$-dependence does serve as a valuable tool in discriminating subleading responses. The exact discrimination power depends on the structure factor in question, those vanishing at $q=0$ being more easily distinguished from the standard response.
While a XENON1T-like experiment would probably only be able to discriminate these structure factors if the corresponding cross section $\sigma_0$ were close to present limits, a XENONnT setting would be able to discriminate an $|\F_\pi|^2$ signal as well, and in addition could identify responses that vanish at $q=0$ for $\sigma_0$ values at or below the expected final sensitivity of XENON1T. In such a case, the most ambitious experimental setting of the planned DARWIN detector would be needed to discriminate an $|\F_\pi|^2$ structure factor, allowing one to identify signals vanishing at $q=0$ even in the case of cross sections suppressed by an additional order of magnitude.  It should be stressed, however, that in extrapolating the detector properties from XENON100 we have most likely underestimated the discrimination power of future experiments, whose sensitivities could be assessed more accurately in the future using refined background and signal models.

Our analysis finds that the lowest discriminating power corresponds to the isovector structure factor $|\F_-^M|^2$, counterpart of the isoscalar Helm form factor, in which case it may be more promising to combine experimental information from different nuclear targets with varied proton-to-neutron ratios. A similar strategy, combined with the search of inelastic scattering~\cite{McCabe:2015eia,Baudis:2013bba}, can be used to identify SD interactions, which can only be observed in nuclear isotopes with an odd number of neutrons or protons.

In addition to quantifying the discrimination power towards subleading nuclear response functions, we have found that in some cases the Helm form factor hypothesis is better distinguished than the ``background only'' scenario from the non-Helm form factor signals. In particular this is the case for all structure factors in this study where there is a dependence on $q^4$. This suggests that a dark matter signal, based on one of these underlying nuclear structure factors, could actually be missed if only the standard SI nuclear response were tested. Our study therefore provides additional motivation to also set limits for other WIMP--nucleus interactions in future experimental analyses. 

\begin{acknowledgments}
We thank Ludwig Rauch for fruitful discussions and insights into the XEPHYR package as well as Ranny Budnik, Hagar Landsman, Daniel Lellouch, and Alessandro Manfredini from the Weizmann Institute of Science for providing the package.
This work was supported in part by the US DOE Grant No.\ DE-FG02-00ER41132,
the ERC Grant No.\ 307986 STRONGINT, the DFG through Grants SFB 1245 and GRK 2149, MEXT (Priority Issue on Post-K computer), and JICFuS.
\end{acknowledgments}


\begin{thebibliography}{99}
	
	\bibitem{Baudis:2016qwx} 
	L.~Baudis,
	%``Dark matter detection,''
	J.\ Phys.\ G {\bf 43}, 044001 (2016).
	%%CITATION = doi:10.1088/0954-3899/43/4/044001;%%
	
	\bibitem{Angloher:2015ewa} 
	G.~Angloher {\it et al.} [CRESST Collaboration],
	%``Results on light dark matter particles with a low-threshold CRESST-II detector,''
	Eur.\ Phys.\ J.\ C {\bf 76}, 25 (2016)
	[arXiv:1509.01515 [astro-ph.CO]].
	%%CITATION = doi:10.1140/epjc/s10052-016-3877-3;%%   
	
	\bibitem{Agnese:2015nto} 
	R.~Agnese {\it et al.} [SuperCDMS Collaboration],
	%``New Results from the Search for Low-Mass Weakly Interacting Massive Particles with the CDMS Low Ionization Threshold Experiment,''
	Phys.\ Rev.\ Lett.\  {\bf 116}, 071301 (2016)
	[arXiv:1509.02448 [astro-ph.CO]].
	%%CITATION = doi:10.1103/PhysRevLett.116.071301;%%
	
	\bibitem{Agnes:2015ftt}
	P.~Agnes {\it et al.} [DarkSide Collaboration],
	%``Results from the first use of low radioactivity argon in a dark matter search,''
	Phys.\ Rev.\ D {\bf 93}, 081101 (2016)  
	[arXiv:1510.00702 [astro-ph.CO]].
	%%CITATION = doi:10.1103/PhysRevD.93.081101;%% 
	
	\bibitem{Amole:2016pye} 
	C.~Amole {\it et al.} [PICO Collaboration],
	%``Improved dark matter search results from PICO-2L Run 2,''
	Phys.\ Rev.\ D {\bf 93}, 061101 (2016)
	[arXiv:1601.03729 [astro-ph.CO]].
	%%CITATION = doi:10.1103/PhysRevD.93.061101;%%
	
	\bibitem{Armengaud:2016cvl} 
	E.~Armengaud {\it et al.} [EDELWEISS Collaboration],
	%``Constraints on low-mass WIMPs from the EDELWEISS-III dark matter search,''
	JCAP {\bf 1605}, 019 (2016)
	[arXiv:1603.05120 [astro-ph.CO]].
	%%CITATION = doi:10.1088/1475-7516/2016/05/019;%%  
	
	\bibitem{Akerib:2016vxi} 
	D.~S.~Akerib {\it et al.} [LUX Collaboration],
	%``Results from a search for dark matter in the complete LUX exposure,''
	Phys.\ Rev.\ Lett.\  {\bf 118}, 021303 (2017)
	[arXiv:1608.07648 [astro-ph.CO]].
	%%CITATION = doi:10.1103/PhysRevLett.118.021303;%%  
	
	\bibitem{Cui:2017nnn}
	X.~Cui {\it et al.} [PandaX-II Collaboration],
	%``Dark Matter Results From 54-Ton-Day Exposure of PandaX-II Experiment,''
	Phys.\ Rev.\ Lett.\  {\bf 119}, 181302 (2017)
	[arXiv:1708.06917 [astro-ph.CO]]
	%%CITATION = doi:10.1103/PhysRevLett.119.181302;%%
	
	\bibitem{Aprile:2017iyp} 
	E.~Aprile {\it et al.} [XENON Collaboration],
	%``First Dark Matter Search Results from the XENON1T Experiment,''
	Phys.\ Rev.\ Lett.\  {\bf 119},  181301 (2017)
	[arXiv:1705.06655 [astro-ph.CO]].
	%%CITATION = doi:10.1103/PhysRevLett.119.181301;%%
	
	\bibitem{Aprile:2015uzo} 
	E.~Aprile {\it et al.} [XENON Collaboration],
	%``Physics reach of the XENON1T dark matter experiment,''
	JCAP {\bf 1604}, 027 (2016)
	[arXiv:1512.07501 [physics.ins-det]].
	%%CITATION = doi:10.1088/1475-7516/2016/04/027;%%
	
	\bibitem{Akerib:2015cja} 
	D.~S.~Akerib {\it et al.} [LZ Collaboration],
	%``LUX-ZEPLIN (LZ) Conceptual Design Report,''
	arXiv:1509.02910 [physics.ins-det].
	%%CITATION = ARXIV:1509.02910;%%
	
	\bibitem{PandaXxT}
	\path{https://static.pandax.sjtu.edu.cn/download/IAC-2016-pandax.pdf}.
	
	\bibitem{Aalbers:2016jon} 
	J.~Aalbers {\it et al.} [DARWIN Collaboration],
	%``DARWIN: towards the ultimate dark matter detector,''
	JCAP {\bf 1611}, 017 (2016)
	[arXiv:1606.07001 [astro-ph.IM]].
	%%CITATION = doi:10.1088/1475-7516/2016/11/017;%%
	
	\bibitem{Akimov:2017ade} 
	D.~Akimov {\it et al.} [COHERENT Collaboration],
	%``Observation of Coherent Elastic Neutrino-Nucleus Scattering,''
	Science {\bf 357}, 1123 (2017)
	[arXiv:1708.01294 [nucl-ex]].
	%%CITATION = doi:10.1126/science.aao0990;%%
	
	\bibitem{Helm:1956zz} 
	R.~H.~Helm,
	%``Inelastic and Elastic Scattering of 187-Mev Electrons from Selected Even-Even Nuclei,''
	Phys.\ Rev.\  {\bf 104}, 1466 (1956).
	%%CITATION = doi:10.1103/PhysRev.104.1466;%%  
	
	\bibitem{Cheung:2012qy} 
	C.~Cheung, L.~J.~Hall, D.~Pinner and J.~T.~Ruderman,
	%``Prospects and Blind Spots for Neutralino Dark Matter,''
	JHEP {\bf 1305}, 100 (2013)
	[arXiv:1211.4873 [hep-ph]].
	%%CITATION = doi:10.1007/JHEP05(2013)100;%%
	
	\bibitem{Huang:2014xua} 
	P.~Huang and C.~E.~M.~Wagner,
	%``Blind Spots for neutralino Dark Matter in the MSSM with an intermediate m_A,''
	Phys.\ Rev.\ D {\bf 90}, 015018 (2014)
	[arXiv:1404.0392 [hep-ph]].
	%%CITATION = doi:10.1103/PhysRevD.90.015018;%%
	
	\bibitem{Crivellin:2015bva} 
	A.~Crivellin, M.~Hoferichter, M.~Procura and L.~C.~Tunstall,
	%``Light stops, blind spots, and isospin violation in the MSSM,''
	JHEP {\bf 1507}, 129 (2015)
	[arXiv:1503.03478 [hep-ph]].
	%%CITATION = doi:10.1007/JHEP07(2015)129;%%
	
	\bibitem{Aprile:2013doa} 
	E.~Aprile {\it et al.} [XENON100 Collaboration],
	%``Limits on spin-dependent WIMP-nucleon cross sections from 225 live days of XENON100 data,''
	Phys.\ Rev.\ Lett.\  {\bf 111}, 021301 (2013)
	[arXiv:1301.6620 [astro-ph.CO]].
	%%CITATION = doi:10.1103/PhysRevLett.111.021301;%%   
	
	\bibitem{Uchida:2014cnn} 
	H.~Uchida {\it et al.} [XMASS-I Collaboration],
	%``Search for inelastic WIMP nucleus scattering on $^{129}Xe$ in data from the XMASS-I experiment,''
	PTEP {\bf 2014}, 063C01 (2014)
	[arXiv:1401.4737 [astro-ph.CO]].
	%%CITATION = doi:10.1093/ptep/ptu064;%%  
	
	\bibitem{Fu:2016ega} 
	C.~Fu {\it et al.} [PandaX-II Collaboration],
	%``Spin-Dependent Weakly-Interacting-Massive-Particle–Nucleon Cross Section Limits from First Data of PandaX-II Experiment,''
	Phys.\ Rev.\ Lett.\  {\bf 118}, 071301 (2017)
	[arXiv:1611.06553 [hep-ex]].
	%%CITATION = doi:10.1103/PhysRevLett.118.071301;%%  
	
	\bibitem{Akerib:2017kat} 
	D.~S.~Akerib {\it et al.} [LUX Collaboration],
	%``Limits on spin-dependent WIMP-nucleon cross section obtained from the complete LUX exposure,''
	Phys.\ Rev.\ Lett.\  {\bf 118}, 251302 (2017)
	[arXiv:1705.03380 [astro-ph.CO]].
	%%CITATION = doi:10.1103/PhysRevLett.118.251302;%% 
	
	\bibitem{Schneck:2015eqa} 
	K.~Schneck {\it et al.} [SuperCDMS Collaboration],
	%``Dark matter effective field theory scattering in direct detection experiments,''
	Phys.\ Rev.\ D {\bf 91}, 092004 (2015)
	[arXiv:1503.03379 [astro-ph.CO]].
	%%CITATION = doi:10.1103/PhysRevD.91.092004;%%  
	
	\bibitem{Aprile:2017aas} 
	E.~Aprile {\it et al.} [XENON Collaboration],
	%``Effective field theory search for high-energy nuclear recoils using the XENON100 dark matter detector,''
	Phys.\ Rev.\ D {\bf 96}, 042004 (2017)
	[arXiv:1705.02614 [astro-ph.CO]].
	%%CITATION = doi:10.1103/PhysRevD.96.042004;%%  
	
	\bibitem{Angloher:2016jsl} 
	G.~Angloher {\it et al.},
	%``Limits on momentum-dependent asymmetric dark matter with CRESST-II,''
	Phys.\ Rev.\ Lett.\  {\bf 117},  021303 (2016)
	[arXiv:1601.04447 [astro-ph.CO]].
	%%CITATION = doi:10.1103/PhysRevLett.117.021303;%%
	
	\bibitem{Epelbaum:2008ga} 
	E.~Epelbaum, H.-W.~Hammer and U.-G.~Mei{\ss}ner,
	%``Modern Theory of Nuclear Forces,''
	Rev.\ Mod.\ Phys.\  {\bf 81}, 1773 (2009)
	[arXiv:0811.1338 [nucl-th]].
	%%CITATION = doi:10.1103/RevModPhys.81.1773;%%  
	
	\bibitem{Machleidt:2011zz} 
	R.~Machleidt and D.~R.~Entem,
	%``Chiral effective field theory and nuclear forces,''
	Phys.\ Rept.\  {\bf 503}, 1 (2011)
	[arXiv:1105.2919 [nucl-th]].
	%%CITATION = doi:10.1016/j.physrep.2011.02.001;%%  
	
	\bibitem{Hammer:2012id} 
	H.-W.~Hammer, A.~Nogga and A.~Schwenk,
	%``Three-body forces: From cold atoms to nuclei,''
	Rev.\ Mod.\ Phys.\ {\bf 85}, 197 (2013)
	[arXiv:1210.4273 [nucl-th]].
	%%CITATION = doi:10.1103/RevModPhys.85.197;%%
	
	\bibitem{Bacca:2014tla} 
	S.~Bacca and S.~Pastore,
	%``Electromagnetic reactions on light nuclei,''
	J.\ Phys.\ G {\bf 41}, 123002 (2014)
	[arXiv:1407.3490 [nucl-th]].
	
	\bibitem{Hoferichter:2016nvd} 
	M.~Hoferichter, P.~Klos, J.~Men\'endez and A.~Schwenk,
	%``Analysis strategies for general spin-independent WIMP-nucleus scattering,''
	Phys.\ Rev.\ D {\bf 94}, 063505 (2016)
	[arXiv:1605.08043 [hep-ph]].
	%%CITATION = doi:10.1103/PhysRevD.94.063505;%%
	
   \bibitem{Rogers:2016jrx} 
   H.~Rogers, D.~G.~Cerde\~no, P.~Cushman, F.~Livet and V.~Mandic,
   %``Multidimensional effective field theory analysis for direct detection of dark matter,''
   Phys.\ Rev.\ D {\bf 95}, 082003 (2017)
   [arXiv:1612.09038 [astro-ph.CO]].
   %%CITATION = doi:10.1103/PhysRevD.95.082003;%%  
	
	\bibitem{Lewin:1995rx} 
	J.~D.~Lewin and P.~F.~Smith,
	%``Review of mathematics, numerical factors, and corrections for dark matter experiments based on elastic nuclear recoil,''
	Astropart.\ Phys.\  {\bf 6}, 87 (1996).
	%%CITATION = doi:10.1016/S0927-6505(96)00047-3;%%  
	
	\bibitem{Kahlhoefer:2016eds} 
  F.~Kahlhoefer and S.~Wild,
  %``Studying generalised dark matter interactions with extended halo-independent methods,''
  JCAP {\bf 1610}, 032 (2016)
  [arXiv:1607.04418 [hep-ph]].
  %%CITATION = doi:10.1088/1475-7516/2016/10/032;%%
	
	\bibitem{Catena:2018ywo} 
  R.~Catena, A.~Ibarra, A.~Rappelt and S.~Wild,
  %``Halo-independent comparison of direct detection experiments in the effective theory of dark matter-nucleon interactions,''
  arXiv:1801.08466 [hep-ph].
  %%CITATION = ARXIV:1801.08466;%%
  
  \bibitem{Krauss:2018pvg} 
  L.~M.~Krauss and J.~L.~Newstead,
  %``Extracting Particle Physics Information from Direct Detection of Dark Matter with Minimal Assumptions,''
  arXiv:1801.08523 [hep-ph].
  %%CITATION = ARXIV:1801.08523;%%
	
	\bibitem{Vietze:2014vsa} 
	L.~Vietze, P.~Klos, J.~Men\'endez, W.~C.~Haxton and A.~Schwenk,
	%``Nuclear structure aspects of spin-independent WIMP scattering off xenon,''
	Phys.\ Rev.\ D {\bf 91}, 043520 (2015)
	[arXiv:1412.6091 [nucl-th]].
	%%CITATION = doi:10.1103/PhysRevD.91.043520;%%   
	
	\bibitem{Cirigliano:2012pq} 
	V.~Cirigliano, M.~L.~Graesser and G.~Ovanesyan,
	%``WIMP-nucleus scattering in chiral effective theory,''
	JHEP {\bf 1210}, 025 (2012)
	[arXiv:1205.2695 [hep-ph]].
	%%CITATION = doi:10.1007/JHEP10(2012)025;%% 
	
	\bibitem{Menendez:2012tm} 
	J.~Men\'endez, D.~Gazit and A.~Schwenk,
	%``Spin-dependent WIMP scattering off nuclei,''
	Phys.\ Rev.\ D {\bf 86}, 103511 (2012)
	[arXiv:1208.1094 [astro-ph.CO]].
	%%CITATION = doi:10.1103/PhysRevD.86.103511;%% 
	
	\bibitem{Klos:2013rwa} 
	P.~Klos, J.~Men\'endez, D.~Gazit and A.~Schwenk,
	%``Large-scale nuclear structure calculations for spin-dependent WIMP scattering with chiral effective field theory currents,''
	Phys.\ Rev.\ D {\bf 88}, 083516 (2013)
	Erratum: [Phys.\ Rev.\ D {\bf 89},  029901 (2014)]
	[arXiv:1304.7684 [nucl-th]].
	%%CITATION = doi:10.1103/PhysRevD.89.029901, 10.1103/PhysRevD.88.083516;%%  
	
	\bibitem{Baudis:2013bba} 
	L.~Baudis, G.~Kessler, P.~Klos, R.~F.~Lang, J.~Men\'endez, S.~Reichard and A.~Schwenk,
	%``Signatures of Dark Matter Scattering Inelastically Off Nuclei,''
	Phys.\ Rev.\ D {\bf 88}, 115014 (2013)
	[arXiv:1309.0825 [astro-ph.CO]].
	%%CITATION = doi:10.1103/PhysRevD.88.115014;%%  
	
	\bibitem{Cirigliano:2013zta} 
	V.~Cirigliano, M.~L.~Graesser, G.~Ovanesyan and I.~M.~Shoemaker,
	%``Shining LUX on Isospin-Violating Dark Matter Beyond Leading Order,''
	Phys.\ Lett.\ B {\bf 739}, 293 (2014)
	[arXiv:1311.5886 [hep-ph]].
	%%CITATION = doi:10.1016/j.physletb.2014.10.058;%%   
	
	\bibitem{Hoferichter:2015ipa} 
	M.~Hoferichter, P.~Klos and A.~Schwenk,
	%``Chiral power counting of one- and two-body currents in direct detection of dark matter,''
	Phys.\ Lett.\ B {\bf 746}, 410 (2015)
	[arXiv:1503.04811 [hep-ph]].
	%%CITATION = doi:10.1016/j.physletb.2015.05.041;%% 
	
	\bibitem{Korber:2017ery} 
	C.~K\"orber, A.~Nogga and J.~de Vries,
	%``First-principle calculations of Dark Matter scattering off light nuclei,''
	Phys.\ Rev.\ C {\bf 96}, 035805 (2017)
	[arXiv:1704.01150 [hep-ph]].
	%%CITATION = ARXIV:1704.01150;%%  
	
	\bibitem{Hoferichter:2017olk} 
	M.~Hoferichter, P.~Klos, J.~Menéndez and A.~Schwenk,
	%``Improved limits for Higgs-portal dark matter from LHC searches,''
	Phys.\ Rev.\ Lett.\  {\bf 119}, 181803 (2017)
	[arXiv:1708.02245 [hep-ph]].
	%%CITATION = doi:10.1103/PhysRevLett.119.181803;%%  
	
	\bibitem{Gazda:2016mrp} 
	D.~Gazda, R.~Catena and C.~Forss\'en,
	%``Ab initio nuclear response functions for dark matter searches,''
	Phys.\ Rev.\ D {\bf 95}, 103011 (2017)
	[arXiv:1612.09165 [hep-ph]].
	%%CITATION = doi:10.1103/PhysRevD.95.103011;%%
	
	\bibitem{Bishara:2016hek} 
	F.~Bishara, J.~Brod, B.~Grinstein and J.~Zupan,
	%``Chiral Effective Theory of Dark Matter Direct Detection,''
	JCAP {\bf 1702}, 009 (2017)
	[arXiv:1611.00368 [hep-ph]].
	%%CITATION = doi:10.1088/1475-7516/2017/02/009;%%     
	
	\bibitem{Bishara:2017pfq} 
	F.~Bishara, J.~Brod, B.~Grinstein and J.~Zupan,
	%``From quarks to nucleons in dark matter direct detection,''
	JHEP {\bf 1711}, 059 (2017)
	[arXiv:1707.06998 [hep-ph]].
	%%CITATION = doi:10.1007/JHEP11(2017)059;%%  
	
	\bibitem{Fan:2010gt}
	J.~Fan, M.~Reece and L.~T.~Wang,
	%``Non-relativistic effective theory of dark matter direct detection,''
	JCAP {\bf 1011}, 042 (2010) 
	[arXiv:1008.1591 [hep-ph]].
	%%CITATION = ARXIV:1008.1591;%%  
	
	\bibitem{Fitzpatrick:2012ix} 
	A.~L.~Fitzpatrick, W.~Haxton, E.~Katz, N.~Lubbers and Y.~Xu,
	%``The Effective Field Theory of Dark Matter Direct Detection,''
	JCAP {\bf 1302}, 004 (2013)
	[arXiv:1203.3542 [hep-ph]].
	%%CITATION = doi:10.1088/1475-7516/2013/02/004;%% 
	
	\bibitem{Anand:2013yka} 
	N.~Anand, A.~L.~Fitzpatrick and W.~C.~Haxton,
	%``Weakly interacting massive particle-nucleus elastic scattering response,''
	Phys.\ Rev.\ C {\bf 89},  065501 (2014)
	[arXiv:1308.6288 [hep-ph]].
	%%CITATION = doi:10.1103/PhysRevC.89.065501;%%  

\bibitem{Co:2012adt} 
G.~Co', V.~De Donno, M.~Anguiano and A.~M.~Lallena,
%``Nuclear proton and neutron distributions in the detection of weak interacting massive particles,''
JCAP {\bf 1211}, 010 (2012)
[arXiv:1211.1787 [nucl-th]].
%%CITATION = doi:10.1088/1475-7516/2012/11/010;%%

\bibitem{Cerdeno:2012ix} 
D.~G.~Cerde\~no, M.~Fornasa, J.-H.~Huh and M.~Peir\'o,
%``Nuclear uncertainties in the spin-dependent structure functions for direct dark matter detection,''
Phys.\ Rev.\ D {\bf 87}, 023512 (2013)
[arXiv:1208.6426 [hep-ph]].
%doi:10.1103/PhysRevD.87.023512

	\bibitem{Aprile:2012vw} 
	E.~Aprile {\it et al.} [XENON100 Collaboration],
	%``Analysis of the XENON100 Dark Matter Search Data,''
	Astropart.\ Phys.\  {\bf 54}, 11 (2014)
	[arXiv:1207.3458 [astro-ph.IM]].
	%%CITATION = doi:10.1016/j.astropartphys.2013.10.002;%%
	
	\bibitem{Aprile:2011dd} 
	E.~Aprile {\it et al.} [XENON100 Collaboration],
	%``The XENON100 Dark Matter Experiment,''
	Astropart.\ Phys.\  {\bf 35}, 573 (2012)
	[arXiv:1107.2155 [astro-ph.IM]].
	%%CITATION = doi:10.1016/j.astropartphys.2012.01.003;%%
	
	\bibitem{Aprile:2012nq}
	E.~Aprile {\it et al.} [XENON100 Collaboration],
	%``Dark Matter Results from 225 Live Days of XENON100 Data,''
	Phys.\ Rev.\ Lett.\  {\bf 109}, 181301 (2012) 
	[arXiv:1207.5988 [astro-ph.CO]].
	%%CITATION = doi:10.1103/PhysRevLett.109.181301;%%  

	\bibitem{Aprile:2009dv} 
    E.~Aprile and T.~Doke,
    %``Liquid Xenon Detectors for Particle Physics and Astrophysics,''
    Rev.\ Mod.\ Phys.\  {\bf 82}, 2053 (2010)
    [arXiv:0910.4956 [physics.ins-det]].
     %%CITATION = doi:10.1103/RevModPhys.82.2053;%%
	
\bibitem{Undagoitia:2015gya} 
  T.~Marrod\'an Undagoitia and L.~Rauch,
  %``Dark matter direct-detection experiments,''
  J.\ Phys.\ G {\bf 43}, 013001 (2016)
  [arXiv:1509.08767 [physics.ins-det]].
  %%CITATION = doi:10.1088/0954-3899/43/1/013001;%%	
	
	\bibitem{Aprile:2014zvw}
	E.~Aprile {\it et al.} [XENON1T Collaboration]
	%"Conceptual design and simulation of a water Cherenkov muon veto for the XENON1T experiment"
	JINST {\bf 9}, P11006 (2014)
	[arXiv:1406.2374 [astro-ph.IM]].
	%%CITATION = ARXIV:1406.2374;%%"

\bibitem{McCabe:2015eia} 
C.~McCabe,
%``Prospects for dark matter detection with inelastic transitions of xenon,''
JCAP {\bf 1605}, 033 (2016)
[arXiv:1512.00460 [hep-ph]].
%%CITATION = doi:10.1088/1475-7516/2016/05/033;%% 		

\end{thebibliography}
\end{document}